\documentclass[a4paper,12pt]{article}
\usepackage[pctex32]{graphics}
\usepackage{amssymb,amsmath}

\begin{document}

\topmargin 0pt \oddsidemargin 0mm
\newcommand{\be}{\begin{equation}}
\newcommand{\ee}{\end{equation}}
\newcommand{\ba}{\begin{eqnarray}}
\newcommand{\ea}{\end{eqnarray}}
\newcommand{\fr}{\frac}

\begin{titlepage}

\vspace{5mm}
\begin{center}
{\Large \bf Thermodynamics of regular black hole}

\vspace{12mm}

{\large   Yun Soo Myung$^{\rm a,}$\footnote{e-mail
 address: ysmyung@inje.ac.kr},
 Yong-Wan Kim $^{\rm a,}$\footnote{e-mail
 address: ywkim65@gmail.com},
and Young-Jai Park$^{\rm b,}$\footnote{e-mail
 address: yjpark@sogang.ac.kr}}
 \\
\vspace{10mm} {\em $^{\rm a}$Institute of Basic Science and School
of Computer Aided Science, \\Inje University, Gimhae 621-749,
Korea
\\} {\em $^{\rm b}$Department of Physics  and Center for Quantum
Spacetime,\\ Sogang University, Seoul 121-742, Korea}
\end{center}

\vspace{5mm}

\centerline{{\bf{Abstract}}}

\vspace{5mm} We investigate thermodynamics  for a magnetically
charged regular black hole (MCRBH), which comes from the action of
general relativity and nonlinear electromagnetics, comparing with
the Reissner-Norstr\"om (RN) black hole in both four and two
dimensions after dimensional reduction. We find that there is no
thermodynamic difference between the regular and RN black holes
for a fixed charge $Q$ in both dimensions. This means that the
condition for either singularity or regularity at the origin of
coordinate does not affect the thermodynamics of black hole.
Furthermore, we describe the near-horizon AdS$_2$ thermodynamics
of the MCRBH with the connection of the Jackiw-Teitelboim theory.
We also identify the near-horizon entropy as the statistical
entropy by using the AdS$_2$/CFT$_1$ correspondence.

\vspace{3mm}

\noindent  Regular black holes; thermodynamics; Jackiw-Teitelboim
theory.
\end{titlepage}

\newpage

\renewcommand{\thefootnote}{\arabic{footnote}}
\setcounter{footnote}{0} \setcounter{page}{2}

\section{Introduction}
Hawking's semiclassical analysis of a black hole radiation
suggests that most information about initial states is shielded
behind event horizon and will not back to asymptotic region far
from an evaporating black hole~\cite{HAW1}. This means that the
unitarity is violated by an evaporating black hole. However, this
conclusion has been debated by many authors for three
decades~\cite{THOO,SUS,PAG}. It is closely related to a long
standing puzzle of  the information loss paradox, which states the
question of whether the formation and subsequent evaporation of a
black hole is unitary. In order to determine the final state of
evaporation process, a more precise treatment including quantum
gravity effects and backreaction is generally required. In the
semiclassical study of Schwarzschild black hole, the temperature
($T^{Sch}_H \propto 1/m$) and the luminosity ($L_{Sch} \propto
1/m^2$) diverge as the mass $m$ of the black hole approaches zero.
This means that the semiclassical approach breaks down for very
light black holes. Furthermore, one has to take into account
backreaction. It was shown that the effect of quantum gravity
could cure this pathological short distance
behavior~\cite{nicol1,nicol2}. Also, if an extremal black hole is
considered as the ground state of regular black hole (RBH), one
may avoid the short distance behavior such as a terminal phase of
evaporation and backreaction.

At present, two leading candidates for quantum gravity are the
string theory and the loop quantum gravity. Interestingly, the
semiclassical analysis of the loop quantum black hole provides a
RBH without singularity  in contrast to the classical
one~\cite{MOD}. Its minimum size $r_c$ is at Planck scale
$\ell_{Pl}$. On the other hand, in the continuing search for
quantum gravity, the black hole thermodynamics may be related to a
future experimental result at the LHC~\cite{LHC1,LHC2,LHC3}. The
causal structures of RBHs are similar to the
Reissner-Nordstr\"{o}m (RN) black hole with the singularity
replaced by de Sitter space-time with curvature radius
$r_0=\sqrt{3/\Lambda}$~\cite{Dymn1,Dymn2,Shanka}. Recently,
several authors have discussed the formation and evaporation
process of a RBH with minimum size $l$~\cite{HAY,REU} induced from
the string theory~\cite{Vene,GM}. The noncommutativity also
provides another RBH with minimum scale $\sqrt{\theta}$ so called
the noncommutative black hole~\cite{nicol1,nicol2,SS,Riz}. Very
recently, we have investigated the thermodynamics and evaporation
process of the noncommutative black hole~\cite{YKP}. It turned out
that the final state of the evaporation process for all RBHs is a
cold Planck-size remnant of extremal black holes with zero
temperature. The connection between their minimum sizes is given
by $r_c \sim r_{0} \sim l \sim \sqrt{\theta} \sim Q \sim
\ell_{Pl}$, where $Q$ is the charge of the RN black hole. We
expect that the thermodynamics of RBHs  is similar to the RN black
hole~\cite{Hisc}, even though the latter has a timelike
singularity~\cite{RNBH}.

In fact, RBHs have been considered, dating back to
Bardeen~\cite{BAR}, for avoiding the curvature singularity beyond
event horizon in black hole physics~\cite{RBH}. Among various RBHs
known to date, intriguing  black holes are obtained  from the
 action of Einstein gravity and nonlinear electrodynamics. The
solutions to the coupled equations were found by Ay\'on--Beato and
Garc\'{\i}a~\cite{9911174} and by Bronnikov~\cite{0006014}. The
latter describes a magnetically charged regular black hole (MCRBH).
 Also its simplicity allows exact treatment such that the
location of the horizons can be expressed in terms of the Lambert
functions~\cite{Lambert}. Moreover, Matyjasek investigated the
 extremal MCRBH with  the near horizon geometry of
$AdS_{2}\times S^{2}$~\cite{mat,bmt}.

On the other hand, 2D dilaton gravity has been used in various
situations as an effective description of 4D gravity after a black
hole in string theory has appeared \cite{witten,MSW}. Hawking
radiation and thermodynamics of this black hole have been analyzed
by several authors \cite{crff,RST,Frolov,FPST,GKV,GMc}. Another 2D
theories, which were originated from the Jackiw-Teitelboim (JT)
theory \cite{JT1,JT2}, have been also studied
\cite{JT-theories1,JT-theories2,JT-theories3}. Although in this JT
theory the curvature is constant and negative, it has a black hole
solution, which implies  the non-trivial thermodynamics
\cite{cacl,AO,LS,CM,KR,Cad}. Moreover, Fabbri {\it et. al.}
\cite{fnn} partially demonstrated the duality of the
thermodynamics between a near-extremal RN black hole and the JT
theory by considering temperature and entropy. Actually, 2D
dilaton gravity approach is the $s$-wave approximation to 4D
gravity~\cite{NO}. Recently, we have studied whether the entropy
function approach~\cite{sen} is  suitable or not by obtaining the
entropy of extremal MCRBH~\cite{mkpe}, and have investigated it in
terms of the attractor mechanism~\cite{mkp}. The key ingredient is
to find a 2D dilaton gravity with dilaton potential~\cite{myp2}.
Note that several authors have recently mentioned how to derive
the desired Bekenstein-Hawking entropy of extremal RBHs from the
generalized entropy formula based on the Wald's Noether charge
formalism~\cite{CC}.

In this paper, we  study thermodynamic properties of the
MCRBH~\cite{mat,mkpe,mkp}. The motivation of studying this MCRBH
is two folds: regularity and nonlinearity. The first issue is the
regularity of the black hole solution. We exactly know the action
for the MCRBH, in contrast to the noncommutative RBH whose action
is unknown. The second one is the nonlinearity. We may introduce
another nonlinear electromagnetics, Born-Infeld action. However,
this action does not lead to a regularity of metric function in
the limit of $r\to 0$ even  though its presence softens the
divergence of curvature scalar.

We observe that  there exists an unstable point at $r_+=r_m$
(known as Davies' point), where the temperature is maximum and the
heat capacity changes from negative infinity to positive infinity.
This Davies' point separates the whole thermodynamic process into
the early stage with positive heat capacity  and the late stage
with negative heat capacity~\cite{Myung:2007qt}. We also confirm
this feature by using the effective 2D dilaton gravity.

\section{Thermodynamic quantities of MCRBH}

 In order to analyze the thermodynamics of the MCRBH,
 let us start with the four-dimensional non-linear action \cite{mat,bmt,mkpe}
\begin{equation}
\label{action} I=\frac{1}{16\pi}\int d^4x \sqrt{-g}[R-{\cal
L}_M(B)],
\end{equation}
where ${\cal L}_M(B)$  is a functional of $B= F_{\mu\nu}F^{\mu\nu} $
defined by
\begin{equation}  \label{lagr}
 {\cal L}_M(B)=B \cosh^{-2}\left[a \left(\frac{B}{2}\right)^{1/4}
 \right].
 \end{equation}
Here the free parameter $a$ will be adjusted to guarantee
regularity at the center. In the limit of $a\to 0$, this action
reduces  to  the Einstein-Maxwell theory having the solution of
the RN black hole.
 First, the tensor field $F_{\mu\nu}$ satisfies equations
\begin{equation}
\label{maxwell1} \nabla _{\mu}\left( \frac{d{\cal L}(B)}{dB}
F^{\mu\nu}\right) =0,
\end{equation}
\begin{equation}
\label{maxwell2} \nabla _{\mu}\,^{\ast }F^{\mu\nu}=0,
\end{equation}
where the asterisk denotes the Hodge duality. Then,
differentiating the action $I$ with respect to the metric tensor
$g_{\mu\nu}$ leads to
\begin{equation} \label{EEQ}
R_{\mu\nu}-\frac{1}{2} g_{\mu\nu}R = 8\pi T_{\mu\nu}
\end{equation}
with the stress-energy tensor
\begin{equation}
T_{\mu\nu}=\frac{1}{4\pi }\left( \frac{d {\cal L}\left( B\right)
}{dB}F_{\rho \mu}F^{\rho}_{\nu}-\frac{1}{4}g_{\mu\nu} {\cal
L}\left( B\right) \right).
\end{equation}

For our purpose, we consider the spherically symmetric metric
\begin{equation}
\label{metric}
 ds^2=-U(r)dt^2+\frac{1}{U(r)}dr^2+ b^2(r) d\Omega^2_2,
\end{equation}
where $b(r)$ plays a role of radius $r$ of the two sphere $S^2$.
 To determine the metric function (\ref{metric}) defined by
\begin{equation}
U(r)\,=\,1\,-\,\frac{2 m(r)}{r},
 \end{equation}
we have to solve the Einstein equation. It leads to the mass
distribution
\begin{equation} \label{quadrature1}
m(r)\,=\,\frac{1}{4}\int^r  {\cal L}[B(r')]r'^{2} dr'\, + C,
\end{equation}
where $C$ is an integration constant.
 In order to determine $m(r)$, from Eq. (\ref{maxwell1})
we choose the purely magnetic configuration by taking $F_{\mu\nu}$
to zero except for $F_{\theta\phi}$ as follows
\begin{equation}
F_{\theta\phi} = Q \sin\theta \to B=\frac{2Q^{2}}{r^{4}},
\end{equation}
where $Q$ is an integration constant related to the magnetic
charge of the solution. Hereafter we assume that $Q>0$ for
simplicity.

Considering the condition for the ADM mass at infinity as
$m(\infty)\,=\,M=C$, the mass distribution takes the form
\begin{equation} \label{mass}
m(r)\,=M-\frac{Q^{3/2}}{2a} \tanh\left(\frac{aQ^{1/2}}{r} \right).
\end{equation}
Moreover, setting $a\,=\,Q^{3/2}/2M$ determines the metric
function (\ref{metric}) completely as
\begin{equation}
U(r)\,=\,1\,-\,\frac{2 M}{r}\left(1\,-\,\tanh\frac{Q^{2}}{2Mr}
\right). \label{Gr}
 \end{equation}
At this stage we note that $U(r)$ is regular ($ U(r)\to 1$) as $r
\to 0$ using $\lim_{r\to 0}\tanh[aQ^{1/2}/r]\sim
1$\footnote{Unless $a=Q^{3/2}/2M$, one could not recover a
regularity at $r\to 0$. Hence, this  choice of $a$ is necessary
and sufficient condition to obtain a regular black hole. One may
consider three-parameter family of $(a,Q,M)$. However, this  is
not the case, which
   could lead to a regular black hole. If this is the case, its
   solution of metric function  has a singularity in the limit of
   $r\to 0$, like a Born-Infeld  black hole.}, in contrast to the RN case ($a\to0$ limit) whose metric
function of $1-2M/r + Q^2/r^2$ diverges as $r^{-2}$ in that limit.
In order to find the horizon from $U(r)=0$, we use the Lambert
functions $W_i (\xi)$ defined by the general formula
$e^{W(\xi)}W(\xi)=\xi$ \cite{Lambert}. Here $W_0(\xi)$ and
$W_{-1}(\xi)$ have real branches. Their values at branch point
$\xi=-1/e$ are the same as $W_{0}(-1/e)=W_{-1}(-1/e)=-1$.
 Here, we set
$W_{0}(1/e) \equiv w_0$ because the value of the principle branch of
the Lambert function at $\xi=1/e=0.368$ plays a role in finding the
location of degenerate horizon of the extremal MCRBH~\cite{mat,mkp}.

Introducing a reduced radial coordinate $x=r/M$ and a charge-to-mass
ratio $q=Q/M$, the condition for the event horizon is given by
\begin{equation}
U(x)\,=\,1\,-\,\frac{2 }{x}\left(1\,-\,\tanh\frac{q^{2}}{2x}
\right)=0. \label{Grn}
 \end{equation}
Here, one finds the outer $x_+$ and inner $x_-$ horizons as
\begin{equation}
x_+(q)=-\frac{q^2}{W_0(-\frac{q^2e^{q^2/4}}{4})-q^2/4},
~~x_-(q)=-\frac{q^2}{W_{-1}(-\frac{q^2e^{q^2/4}}{4})-q^2/4}.
\end{equation}
For $q=q_{e}=2\sqrt{w_0}$, the two horizons $x_+$ and $x_-$ merge
into a degenerate event horizon  at
\begin{equation} \label{extsol}
x_{e}=\frac{4q^2_{e}}{4+ q^2_{e}}=\frac{4w_0}{1+w_0},
\end{equation}
where we have used the relation of
$(q_{e}^2/4)e^{q_{e}^2/4}=1/e=w_0e^{w_0}$. That is, the degenerate
horizon numerically appears  at $(q_{e}=1.056, x_{e}=0.872)$ when
$x_+=x_-=x_e$. Formally, Eq. (\ref{extsol}) comes from the
extremal condition of $U'(x)=0$. We have an ambiguity to determine
the mass $M_e$ of the extremal MCRBH. For simplicity, we choose
$M_e=1$, and then $Q_e=M_eq_{e}=q_e$. On the other hand, for
$q>q_{e}$ there is no horizon while two horizons appear for
$q<q_e$. For comparison, we note the difference between
$M_e=Q/1.056$ for the extremal MCRBH and $M_e=Q$ for the extremal
RN black hole.
\begin{figure}[t!]
   \centering
   \includegraphics{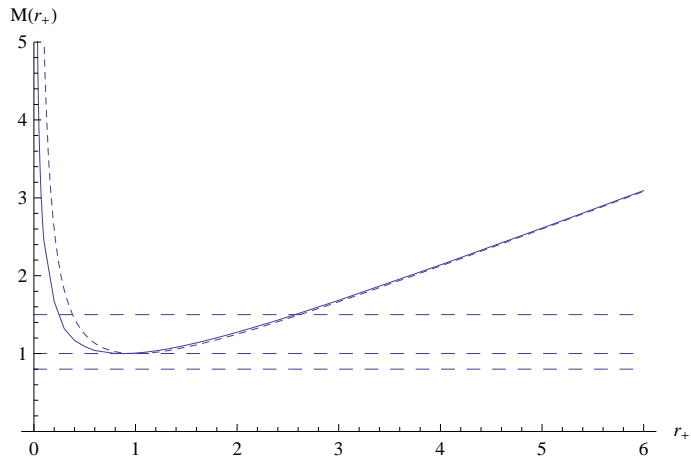}
\caption{Graph of the horizon  mass $M$ versus the horizon radius
$r_\pm$ as the solution to $U(r_\pm)=0$ with a fixed $Q=Q_e$. The
solid (dashed) curve describes the MCRBH (RN). For $M=M_e$, the
degenerate event horizon is located at $r_e=0.872$, while at
$r_e=1$ for the RN black hole. Three horizontal lines are for
$M=1.5,~1,~0.8$.} \label{fig1}
\end{figure}

From the condition $U(r_\pm)=0$ for the horizons, one finds the
mass as a function of horizon radius $r_\pm$ as
\begin{equation} \label{mfun}
M(r_\pm)\equiv
M_\pm=\frac{r_\pm}{2\Big[1-\tanh\Big(\frac{Q^2}{2M_\pm
r_\pm}\Big)\Big]},
\end{equation}
which is obviously a nonlinear relation between $M_\pm$ and
$r_\pm$ due to preserving the regularity. Actually, the
nonlinearity makes the thermodynamic analysis difficult. In order
to see the relation, we plot the horizon mass $M$ as a function of
the horizon radius $r=r_\pm$ for a fixed $Q=Q_e$ numerically in
Fig. 1. The degenerate event horizon locates at $r=r_e=0.872$,
where the minimum mass $M(r_e)=M_e$ appears from Eq.(\ref{mfun}).
Note that for $M(0.8) <M_e$, there is no horizon, which means that
any solution to Eq. (\ref{mfun}) does not exist, whereas for
$M(1.5)>M_e$ one has two horizons: the inner $r=r_-$ and outer
$r=r_+$ horizons. For a large $r>r_e$, we have the Schwarzschild
relation $M=r_+/2$. This picture is similar to the case proposed
by Hayward~\cite{HAY,YKP} for a RBH.

Hereafter we consider the outer horizon $r=r_+$ only because we
are interested in the thermodynamic analysis of the RBH. For our
purpose, let us define the Bekenstein-Hawking entropy for the
MCRBH as
\begin{equation} \label{BHRBH}
S_{BH}= \pi r^2_{+}.
\end{equation}

\begin{figure}[t!]
   \centering
   \includegraphics{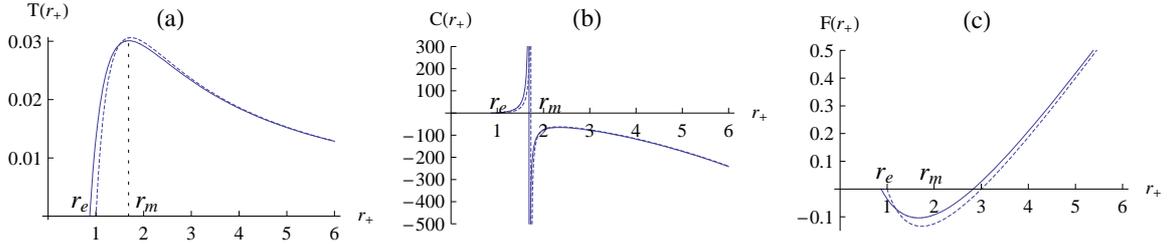}
\caption{Three graphs for temperature, heat capacity, and
free energy with a fixed $Q=Q_e$. The solid (dashed) curve denotes
MCRBH (RN). The near-horizon thermal phase takes place
for $r_e < r_+ <r_m$, the Schwarzschild phase is for $r_+>r_m$. (a)
Graph for the temperature $T$ having the maximum value at $r_+=r_m$.
 (b) Graph
for the heat capacity $C$ showing the blow-up at $r_+=r_m$. The
near-horizon thermodynamics takes the positive heat capacity $C >
0$, while the Schwarzschild phase has the negative heat capacity
$C<0$. (c) Plot of the  free energy $F$.} \label{fig2}
\end{figure}

The black hole temperature can be calculated
 to be
\begin{eqnarray} \label{2eq4}
 T(r_+)&=& \frac{1}{4\pi}~\Bigg[\frac{dU}{dr}\Bigg]_{r=r_+}
   \nonumber\\
  &=&\frac{1}{4\pi}\left[\frac{1}{r_+} + \frac{Q^2}{4M^2_+ r_+}\left(1-\frac{4M_+}{r_+}\right)\right].
\end{eqnarray}
Note that one recovers the Hawking temperature $T^{Sch}_H \propto
r^{-1}_+$ of the Schwarzschild black hole for $r_+>r_m$ with the
Davies' point $r_m$, where the Hawking temperature reaches to the
maximum value at $r_+=r_m$ as shown in Fig.~\ref{fig2}. It is
important to investigate what happens as $r_+ \to 0$. In the
Schwarzschild case, $T^{Sch}_H$ diverges and this puts the limit
on the validity of the evaporation process via the Hawking
radiation. Against this scenario, the temperature $T$ falls down
to zero at $r_+=r_e$\footnote{ The extremal black hole seems to be
controversial because the entropy is non-zero ($S_e=\pi r_e^2$),
while its temperature is zero. This is a long-standing problem for
the extremal black hole. However, our guideline is that the
first-law of thermodynamics should hold even for the extremal
configuration and thus, it remains one of equilibria. In this
case, we prove that the first-law is satisfied as $dM=TdS=0$ at
$M=M_e$. Hence the above case is compatible with the first-law of
black hole thermodynamics.} even where the extremal black hole
appears as shown in Fig. 2(a).

As is depicted  in Fig. 2(a), the temperature of the MCRBH grows until
it reaches to the maximum value $T_m \simeq 0.03$ at $r_+=r_m \simeq
1.689~(M=M_m=1.166)$. As a result, the thermodynamics process is
split into the right branch of $r_m < r_+< \infty$ called the
Schwarzschild phase and the left branch of $ r_e \le r_+<r_m$ called
the near-horizon thermal phase. In particular, one has the extremal
black hole at $r_+=r_e$ with $T(r_e)=0$. In the region of $r<r_e$,
there is no black hole for $M<M_e$ and thus the temperature cannot
be defined. For $M>M_e$, we have the inner horizon at $r=r_-$ inside
the outer horizon, but an observer at infinity does not recognize
the presence of this horizon. Hence, we regard this region as the
forbidden region in view of thermodynamic aspects.

In order to check the thermal stability of the MCRBH, we have to know
the heat capacity~\cite{BR}. Its heat capacity
$C=\frac{dM(r_+)}{dT(r_+)}\mid_{Q}$ is calculated in appendix and
given by
\begin{equation} \label{2eq6}
C(r_+)= \frac{16\pi M^3_+ r_+ (4M^2_+ r_+ -4M_+ Q^2 + Q^2 r_+)}
{16M^2_+ Q^2 + 32M^3_+ Q^2 r_+ - r^2_+(4M^2_+ + Q^2)^2},
\end{equation}
where its variation  is plotted in Fig. 2(b). Here, we find a stable
region of  $C>0$, which represents the near-horizon thermodynamics.
We observe that a thermodynamically unstable region ($C< 0$) appears
for $r_+>r_m$ like the Schwarzschild black hole. We note that
$C(r_e)=0$ for the extremal black hole.

It is appropriate to comment on the value of $r_m=1.689$ at which
not only the Hawking temperature reaches to the maximum value, but
also the specific heat blows up. In order to find the position
$r_+=r_m$ correctly, one has to include the variation of the mass
function (\ref{mfun}), as discussed in the appendix. Its value is
shifted toward the inside of the black hole, when compared with the
radius, $r^{RN}_m=1.732$, of the RN black hole. This means that the
MCRBH could be thermodynamically stable in the more restricted region
than the RN black hole's one. This is of course caused by the
nonlinear mass function (\ref{mfun}).

Finally, we may discuss a possible phase transition near $T=0$
by introducing the Helmholtz free energy~\cite{CEJM} as
\begin{equation} \label{2eq7}
F(r_+)=M(r_+)-M_e-T(r_+)S_{BH}(r_+).
\end{equation}
Its graph is shown in Fig. 2(c). The Helmholtz free energy is zero ($F = 0$) at
$r_+=r_e$, as  $F^{RN}_{min} (r_e = 1) = 0$ for the RN black
hole. Both are monotonically decreasing functions of $r_e \le r_+<r_m$. For
$r_+ > r_m$, one finds the Schwarzschild's free energy of $r_+
/4$.

As is observed from Fig. 2, we split the whole thermal process into
the near-horizon thermal and the Schwarzschild phase. The former is
characterized by the increasing temperature and positive heat
capacity, while the latter is determined by the decreasing
temperature and negative heat capacity. We note that the
near-horizon thermodynamics sharply contrasts to the conventional
thermodynamics of the Schwarzschild black hole. Hence it is very
important to explore  thermodynamics of the MCRBH by using the other
approach.

\section{2D dilaton gravity approach of MCRBH}
Various black holes in four dimensions have been widely studied
through the dimensional reduction. Recently, its interest has
increased as an example of AdS$_2$ arising as a near-horizon
geometry.  Very recently, we have shown that the 2D dilaton
gravity approach provides all thermodynamic quantities of
spherically symmetric RBHs in a simple way~\cite{myp2}. In this
section, we shall explicitly show that the 4D MCRBH is equivalent
to a 2D dilaton gravity.

After the dimensional reduction by integrating the action in Eq.
(\ref{action}) over $S^2$, the reduced effective action in two
dimensions is obtained as~\cite{fnn}
\begin{equation} \label{2Daction}
I^{(2)}=\int d^2x\sqrt{-g}
     \left[\frac{1}{4}(b^2 R_2+2g^{\mu\nu}\nabla_\mu b\nabla_\nu b+2)-b^2{\cal
     L}_M\right].
\end{equation}
It is convenient to eliminate the kinetic term by using the
conformal transformation
\begin{equation}
\bar{g}_{\mu\nu}=\sqrt{\phi}~g_{\mu\nu},~\phi=\frac{b^2(r)}{4}.
\end{equation}
Then, we obtain  the action of 2D dilation gravity with $G_2=1/2$
\cite{JT1,JT2}
\begin{equation}\label{JTA}
\bar{I}_{MCRBH}=\int d^2x
\sqrt{-\bar{g}}\left[\phi\bar{R}_2+V(\phi)\right].
\end{equation}
Here, the Ricci scalar and the dilaton potential are
\begin{eqnarray}
\bar{R}_2&=&-\frac{U''}{\sqrt{\phi}},\\
V(\phi)&=&\frac{1}{2\sqrt{\phi}}-
\frac{Q^2}{8\phi^{3/2}}\cosh^{-2}\left[\frac{Q^2}{4M\sqrt{\phi}}\right],
\end{eqnarray}
respectively.  The two equations of motion are
\begin{eqnarray}
\label{tee33} \nabla^2\phi&=& V(\phi), \\
\bar{R}_2&=&-V'(\phi) \label{tee34},
\end{eqnarray}
where the derivative of $V'(\phi)$ takes the form
\begin{eqnarray}
\label{depot}V'(\phi)&=& -\frac{1}{4\phi^{3/2}}
+\frac{3Q^2}{16\phi^{5/2}}\cosh^{-2}\Big[\frac{Q^2}{4M\sqrt{\phi}}\Big] \\
\nonumber
&-&\frac{Q^4}{32M\phi^{3}}\cosh^{-3}\Big[\frac{Q^2}{4M\sqrt{\phi}}\Big]\sinh\Big[\frac{Q^2}{4M\sqrt{\phi}}\Big].
\end{eqnarray}
By choosing a conformal gauge of $\bar{g}_{tx}=0$ \cite{GKL,cfnn},
we obtain the general solution to Eqs. (\ref{tee33}) and
(\ref{tee34}) as
\begin{eqnarray}
\frac{d\phi}{dx}&=&2(J(\phi)-{\cal C}), \\
\label{pot-met}
 ds^2&=&-(J(\phi)-{\cal C})dt^2+\frac{dx^2}{J(\phi)-{\cal C}},
\end{eqnarray}
where $J(\phi)$ is  the integration of $V(\phi)$
\begin{equation}
J(\phi)=\int^{\phi}V(\tilde{\phi})d\tilde{\phi}=\sqrt{\phi}+M
\tanh\Big(\frac{Q^2}{4M\sqrt{\phi}}\Big).
\end{equation}
Here, ${\cal C}$ is a coordinate-invariant constant of the
integration, which is identified with the mass $M$ of the MCRBH.
$J(\phi)$, $V(\phi)$, and $V'(\phi)$ are depicted in Fig. 3.
\begin{figure}[t!]
   \centering
   \includegraphics{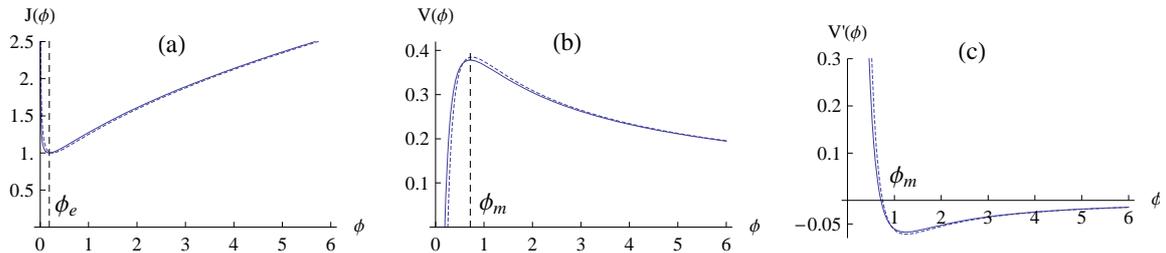}
\caption{Three graphs for $J(\phi),~V(\phi)$, and $V'(\phi)$ with
$Q_e=1.056$. The solid (dashed) curve describes the MCRBH (RN). $J(\phi)$ has a minimum at $\phi_e=0.189$, $V(\phi_m)$
has a maximum value at $\phi_m=0.714$, and $V'(\phi_m)=0$, while for
the extremal RN black holes those are at $\phi^{RN}_e=0.25$ and
$\phi^{RN}_m=0.75$. } \label{figpo}
\end{figure}

We note here the important connection between $J(\phi)$ and the
metric function $U(r(\phi))$ with $r=2\sqrt{\phi}$:
$\sqrt{\phi}~U(\phi)=J(\phi)-M$. A necessary condition that a 2D
dilaton gravity admits an extremal MCRBH is that there exists at
least one curve of $\phi=\phi_e={\rm const}$ such that
$J(\phi_e)=M_e$. In addition, $J(\phi)$ is monotonic in a
neighborhood of $\phi_e=r_e^2/4$ with $J'(\phi_e)=V(\phi_e)=0$ and
$J''(\phi_e)=V'(\phi_e)\not=0$. The initial condition of the
AdS$_2$-horizon $J(\phi_{\pm})=M_\pm$ implies the outer ($\phi_+$)
and inner ($\phi_-$) horizons, which satisfy
\begin{equation} \label{EHJT}
1-\frac{M_\pm}{\sqrt{\phi_\pm}} \Big[1-\tanh
\Big(\frac{Q^2}{4M_\pm \sqrt{\phi_\pm}}\Big)\Big]=0 \to
U(\phi_\pm)=0.
\end{equation}
This  is precisely the definition of the mass function $M_\pm$ in
Eq. (\ref{mfun}). Further conditions on the minimum value
$J(\phi_e)=M_e$ in favor of its extremal configuration imply
\begin{equation} \label{cforbh}
U'(\phi_e)=0,~ U''(\phi_e)\not=0,
\end{equation}
which are the conditions for the degenerate horizon
$r=r_e(Q_e=q_e)$. Hence, for $Q_e=q_e=2\sqrt{w_0}$ and $M_e=1$, we
find the location of the degenerate horizon $r_e=x_e=w_0/(1+w_0)$.
Here, we have an AdS$_2$ spacetime  with negative constant curvature
\begin{equation}
\bar{R}_2|_{r=r_e}=-\frac{2h}{\sqrt{\phi_e}}=
-\frac{1}{\sqrt{\phi_e}}U''(r_e)=-\frac{(1+\omega_0)^4}{32M_e^3\omega^3_0}=-V'(\phi_e).
\end{equation}
There exists an unstable point of $\phi=\phi_m=0.714$, which
satisfies $J'(\phi_m)=V(\phi_m),~ J''(\phi_m)=V'(\phi_m)=0$.

Then, all thermodynamic quantities found in the previous section can
be explicitly expressed in terms of the dilaton $\phi_+$, the
dilaton potential $\widetilde{V}(\phi_+)$, its integration
$\widetilde{J}(\phi_+)$, and its derivative $\widetilde{V}'(\phi_+)$
as
 \begin{eqnarray} \label{phitds}
 &{}&S_{BH}(\phi_+)= 4\pi \phi_+,~~~~~~~T_{H}(\phi_+)=\frac{\widetilde{V}(\phi_+)}{4\pi}, \nonumber \\
 &{}&C(\phi_+) = 4\pi
 \frac{\widetilde{V}(\phi_+)}{\widetilde{V}'(\phi_+)},~~~
F(\phi_+)=\widetilde{J}(\phi_+)-J(\phi_e)-\phi_+
\widetilde{V}(\phi_+),
\end{eqnarray}
where
\begin{eqnarray} \label{vp}
\widetilde{V}(\phi_+)&=&\frac{1}{2\sqrt{\phi_+}}-
\frac{Q^2}{8\phi^{3/2}_+}\cosh^{-2}\left[\frac{Q^2}{4M_+\sqrt{\phi_+}}\right],\nonumber \\
\widetilde{J}(\phi_+)&=&\sqrt{\phi_+}+M_+ \tanh\Big(\frac{Q^2}{4M_+\sqrt{\phi_+}}\Big),\nonumber \\
\widetilde{V}'(\phi_+)&=&\frac{16\pi M^3_+ \phi_+
(4M^2_+\sqrt{\phi_+}-2M_+
Q^2+Q^2\sqrt{\phi_+})}{8M^2_+Q^2\sqrt{\phi_+}-8M^3_+\phi_++2M_+Q^4-Q^4
\sqrt{\phi_+}-2M_+Q^2\phi_+ }.
\end{eqnarray}
We note  the difference between $V, J, V'$ and $\widetilde{V},
\widetilde{J}, \widetilde{V}'$. The former quantities are obtained
by considering the mass $M$ as a constant, while the latter are
obtained by considering the mass $M(r_+)$ as a function of $r_+$.
Hence, for thermodynamic calculations we have to use the tilled
variables $\widetilde{V}, \widetilde{J}$, and $\widetilde{V}'$.

\begin{figure*}[t!]
   \centering
   \includegraphics{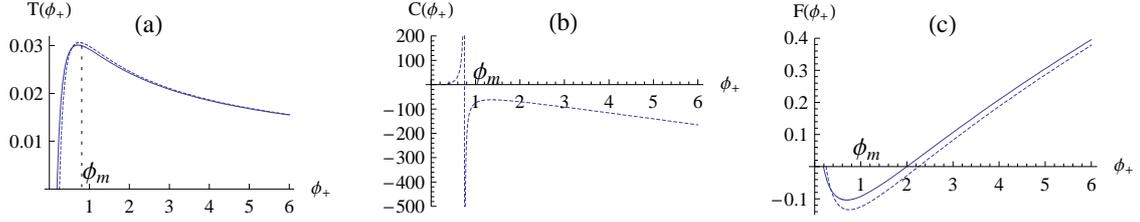}
\caption{Graphs for the thermodynamic quantities as the functions of
$\phi_+$. Here, $4\pi\phi_+$ plays the role of the entropy. The
solid (dashed) curve represents the MCRBH (RN) with
$Q_e = 1.056~(Q_e = 1)$. The regions in $\phi_e\leq\phi_+<\phi_m$
represent the JT phase corresponding to the near-horizon geometry of
the MCRBH.} \label{figth}
\end{figure*}
In Fig. 4, we have the corresponding dual graphs, which are nearly
the same as in Fig. 2. For $\phi_e<\phi<\phi_m$, we have the JT
phase, whereas for $\phi>\phi_m$, we have the Schwarzschild phase.
At the extremal point with $M_e=1$ and $Q_e=1.056$, we have
$T_{H}=0,~C=0$, and $F=0$, which are determined by $V(\phi_e)=0$. On
the other hand, at the maximum point ($M=M_m$), one has
$T_{H}=T_m,~C=\pm \infty$, which are fixed by $V'(\phi_m)=0$.

\section{Near-horizon thermodynamics of extremal MCRBH}
It is a nontrivial task to directly find the near-horizon
thermodynamics from the full thermodynamic quantities because
there exists a nonlinear dependence between the mass $M$ and the
horizon radius $r_+$ in the near-horizon geometry of the 4D
extremal MCRBH. Instead, we use the 2D dilaton gravity because it
was proved that the near-horizon thermodynamics could be
effectively described by the corresponding JT theory for the RN
black hole~\cite{JT1,JT2}. In order to find the AdS$_2$ gravity of
the JT theory, we consider perturbation around the degenerate
event horizon as
\begin{eqnarray}
J(\phi)&=&J(\phi_e)+J'(\phi_e)\varphi+\frac{J''(\phi_e)}{2}\varphi^2
 = M_e+\frac{V'(\phi_e)}{2} \varphi^2,\\
       M&=&M_e[1+ k \alpha^2]\equiv M_e+\Delta M \end{eqnarray}
with $\varphi=\phi-\phi_e$. Although $\widetilde{V}, \widetilde{J}$,
and $\widetilde{V}'$ should be used for thermodynamic calculation,
here we use $V, J$, and $V'$, respectively, for perturbation. This
is because in the near-horizon one has $V\approx \widetilde{V}$,
$J\approx \widetilde{J}$, and $V'\approx \widetilde{V}'$. That is,
$\frac{dM_+}{dr_+} \approx 0$ near the degenerate horizon.

Introducing the new coordinates
\begin{equation} \label{coord}
\tilde{t}=\alpha t,~\tilde{x}=\frac{x-x_e}{\alpha},
\end{equation}
the perturbed dilaton and the metric are given by
\begin{eqnarray}
\varphi&=& \alpha \tilde{x}, \\
 ds^2_{AdS_2}&=&-\Big[\frac{V'(\phi_e)}{2}\tilde{x}^2-kM_e\Big]d\tilde{t}^2+\frac{d\tilde{x}^2}
 {\Big[\frac{V'(\phi_e)}{2}\tilde{x}^2-kM_e\Big]},
 \label{metricads}
\end{eqnarray}
which show a locally AdS$_2$ spacetime. If $k=0$, it is a global
AdS$_2$ spacetime. Moreover, the mass deviation $\Delta M$ is the
conserved parameter of the JT theory~\cite{cfnn}
\begin{equation}
\Delta M=\frac{V'(\phi_0)}{2}\varphi^2-|\nabla\varphi|^2.
\end{equation}
Thus, the JT theory describes both the extremal ($\Delta M=0$) and
the near-extremal ($\Delta M\not=0$) MCRBHs.

Now, we are in a position to derive the near-horizon AdS$_2$
thermodynamic quantities  from the JT theory. From the null
condition of the metric function in Eq. (\ref{metricads}), we have
the positive root
\begin{equation}
\tilde{x}_+=\sqrt{\frac{2kM_e}{V'(\phi_e)}},~~\varphi_+=\sqrt{\frac{2
\Delta M}{V'(\phi_e)}}.
\end{equation}
Then, the JT entropy and temperature are given by
\begin{eqnarray}
S_{JT}&=&4\pi\varphi_+=4\pi\sqrt{\frac{2
\Delta M}{V'(\phi_e)}}, \\
 T_{JT}&=&\frac{V'(\phi_e)}{2\pi}\frac{\varphi_+}{2}=\frac{1}{4\pi}\sqrt{2
 V'(\phi_e)\Delta M}.
\end{eqnarray}
Furthermore, we may have the JT heat capacity and the free energy
\begin{eqnarray}
C_{JT}&=&4 \pi \varphi_+=4\pi \sqrt{\frac{2\Delta M}{V'(\phi_e)}}, \\
 F_{JT}&=&-\phi_e V'(\phi_e)\varphi_+=
 -\frac{(M_ex_e)^2}{4} \sqrt{2V'(\phi_e)\Delta M}.
\end{eqnarray}
Note that $S_{JT}=C_{JT}$ as the case of the RN black hole as
shown in Ref. \cite{myp2}. Finally, all thermodynamic quantities
take the following forms in the near-horizon region:
\begin{eqnarray}
S^{NH}_{BH}&=&S_{BH}(M_e)+S_{JT}=\pi M_e^2 + 4\pi\sqrt{\frac{2
\Delta M}{V'(\phi_e)}}, \\
 T^{NH}_{H}&=&T_{H}(M_e)+T_{JT}=\frac{\sqrt{2
 V'(\phi_e)\Delta M}}{4\pi},\\
C^{NH}&=& C(M_e)+C_{JT}=4\pi \sqrt{\frac{2\Delta M}{V'(\phi_e)}}, \\
 F^{NH}&=&F(M_e)+F_{JT}=-\frac{(M_ex_e)^2}{4} \sqrt{2V'(\phi_e)\Delta M}.
\end{eqnarray}
\begin{figure}[t!]
   \centering
   \includegraphics{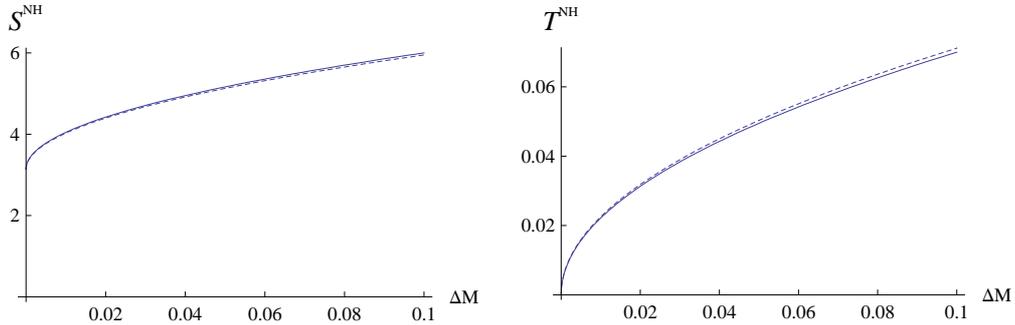}
   \caption{ Plot of the near-horizon (NH) entropy and temperature as functions
   of $\Delta M$ for $r_e \le r_+<r_m$.
   Both are proportional to  $\sqrt{\Delta M}$.} \label{fig5}
\end{figure}
\begin{figure}[t!]
   \centering
   \includegraphics{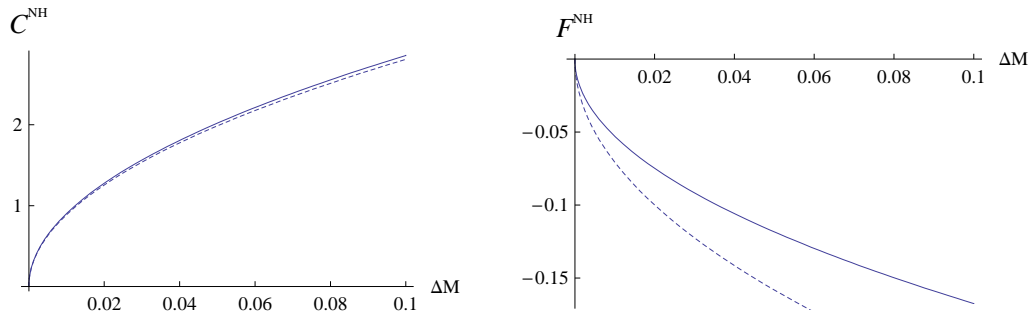}
   \caption{ Plot of the near-horizon heat capacity and  free energy as functions
   of $\Delta M$ for $r_e \le r_+<r_m$.} \label{fig6}
\end{figure}
From Figs. 5 and 6, one finds that there is no thermodynamic
difference between the MCRBH and RN black hole.

\section{AdS$_2$/CFT$_1$ correspondence for entropy}
In this section we interpret the JT entropy $S_{JT}$  to be a
statistical entropy by using AdS$_2$/CFT$_1$ correspondence
according to the previous work~\cite{nn}. This correspondence is
available because of the near horizon isometry of SO(2,1) and an
infinitely long throat of the AdS$_2$ spacetime near the extremal
black hole. If the $\tilde{t}$ in the AdS$_2$ plays the role of a
null coordinate, one may impose asymptotic symmetries on the
boundary (mimicking the analysis of the 3D gravity) as
\begin{eqnarray}
g_{\tilde{t}\tilde{t}}&=-&\frac{\bar{R}_e}{2}\tilde{x}^2+ \gamma_{\tilde{t}\tilde{t}}+\cdots, \\
g_{\tilde{t}\tilde{x}}&=&\frac{\gamma_{\tilde{t}\tilde{x}}}{\tilde{x}^3}+\cdots,\\
g_{\tilde{x}\tilde{x}}&=&\frac{2}{\bar{R}_e}\frac{1}{\tilde{x}^2}+\frac{\gamma_{\tilde{x}\tilde{x}}}{\tilde{x}^4}
+\cdots
\end{eqnarray}
with $\bar{R}_e\equiv \bar{R}_2|_{r=r_e}=-V'(\phi_e)$. Choosing
the boundary conformal gauge with $\gamma_{\tilde{t}\tilde{x}}=0$,
the charges can be derived easily. The infinitesimal
diffeomorphisms $\zeta^a(\tilde{x},\tilde{t})$ preserving the
above boundary conditions are
$\zeta^{\tilde{t}}=\epsilon(\tilde{t}),~\zeta^{\tilde{x}}=-\tilde{x}\epsilon'(\tilde{t})$.
Its action on the 2D gravity  in Eq.(\ref{2Daction}) induces the
following transformation for the function
$\Theta_{\tilde{t}\tilde{t}}=\kappa\Big[\gamma_{\tilde{t}\tilde{t}}-(\bar{R}_e/2)^2\gamma_{\tilde{x}\tilde{x}}/2\Big]$:
\begin{equation}\label{1Dvira}
\delta_{\epsilon}\Theta_{\tilde{t}\tilde{t}}
=\epsilon(\tilde{t})\Theta'_{\tilde{t}\tilde{t}}+2\Theta_{\tilde{t}\tilde{t}}\epsilon'(\tilde{t})
+\frac{2\kappa}{\bar{R}_e}\epsilon'''(\tilde{t}).
\end{equation}
$\Theta_{\tilde{t}\tilde{t}}$ behaves as the chiral component of the
stress tensor of a boundary conformal field theory. To find its
central charge, we have to know the coefficient $\kappa$ in
Eq.(\ref{1Dvira}). For this purpose, we construct the full
Hamiltonian $H=H_0+K$, where $K$ is a boundary term to have
well-defined variational derivatives. This is determined as
$K(\epsilon)=\epsilon(\tilde{t})2\alpha\Big[\gamma_{\tilde{t}\tilde{t}}-
(\bar{R}_e/2)^2\gamma_{\tilde{x}\tilde{x}}/2\Big]$ with
$\kappa=2\alpha$. Assuming a periodicity of $2\pi \beta $ in
$\tilde{t}$~\cite{nn}, we find the central charge and its Virasoro
generator
\begin{equation}
c=-\frac{48 \alpha}{\bar{R}_e\beta},~~L^R_0=M_ek\alpha\beta.
\end{equation}
Using the Cardy-formula for the right movers, one has the desired
statistical entropy as follows
\begin{equation}
S_{st}^{CFT_1}=2\pi \sqrt{\frac{cL_0^R}{6}}=2 \pi \sqrt{\frac{8
M_e k\alpha^2}{-\bar{R}_e}}=4\pi \sqrt{\frac{2 \Delta
M}{V'(\phi_e)}}=S_{JT}.
\end{equation}
This statistical entropy accounts for the microscopic excitations
around the extremal macroscopic state of the MCRBH.

\section{Discussions}
There are a few of approaches to understanding a magnetically
charged regular black hole (MCRBH). However, it remains a
nontrivial task to understand its full thermodynamic behaviors
because this MCRBH was constructed from the combination of
Einstein gravity and nonlinear electromagnetics. In this work, we
have explored the thermodynamics of the MCRBH completely. Here,
the extremal MCRBH is determined by zero temperature $(T=0$), zero
heat capacity ($C=0$), and zero free energy $(F=0)$. We have also
found an important point where the temperature is maximum, the
heat capacity changes from positive infinity to negative infinity.
This point separates the whole thermodynamic process into the
near-horizon phase  with positive heat capacity and the
Schwarzschild phase with negative heat capacity.  The former
represents the near-horizon AdS$_2$ thermodynamics of the extremal
MCRBH, which is characterized by the increasing temperature,
positive heat capacity, and decreasing
 free energy. We have also reexamined the thermodynamics of the MCRBH
by using the 2D dilaton gravity and its near-horizon
thermodynamics by introducing the Jackiw-Teitelboim theory of
AdS$_2$-gravity. All thermodynamic behaviors of the MCRBH are
similar to those of the singular RN black hole. This means that an
observer at infinity does unlikely distinguish between the regular
and the singular black holes.

Concerning a possible phase transition, one expects that a phase
transition occurs near $T=0$, from the extremal MCRBH to the
non-extremal MCRBH. However, in order to study the presumed phase
transition, we have to introduce the negative cosmological
constant because the free energy is  positive for large $r_+$
\cite{BD07}. Having the AdS-RBH, one may find the negative free
energy for large $r_+$. Then, we may discuss the phase transition
from the extremal MCRBH at $r_+=r_e$ to a large MCRBH with $r_+
\gg r_e$ in AdS spacetime, similar to the Hawking-Page transition
from the thermal AdS spacetime at $r_+=0$ to a large black
hole~\cite{Hawking1,Hawking2}.

In conclusion, we have shown that the thermodynamic behaviors of
the MCRBH without singularity is the nearly same as those of the
RN black hole with singularity. This is because the temperature in
Fig. 2(a), the heat capacity in Fig. 2(b), and the free energy in
Fig. 2(c) show the nearly same behaviors, regardless of
singularity and regularity at the origin.

\section*{Acknowledgement}
Two of us (Y.S. Myung and Y.-J. Park) were supported by the Science
Research Center Program of the Korea Science and Engineering
Foundation through the Center for Quantum Spacetime of Sogang
University with grant number R11-2005-021. Y.-W. Kim was supported
by the Korea Research Foundation Grant funded by Korea Government
(MOEHRD): KRF-2007-359-C00007.

\section*{Appendix: Proofs of Eqs. (\ref{2eq6}) and (\ref{vp}) }

In this appendix, we will show how to get the concrete form of the
specific heats for the two approaches. In the definition of the
specific heat as
\begin{eqnarray}
C=\left(\frac{dM}{dT}\right)_Q &=& \frac{dM(r_+)}{dr_+}\frac{dr_+}{dT(r_+)} \label{thermo-cp}\\
               &=& \frac{dM(\phi_+)}{d\phi_+}\frac{d\phi_+}{dT(r_+)},
               \label{pot-cp}
\end{eqnarray}
the derivatives of the mass functions $M(r_+)$ $(M(\phi_+))$ with
$r_+$ ($\phi_+$) can be easily obtained from the metric function
$U(r_+)=0$ in Eq.(\ref{Gr}) and $J(\phi)-M=0$ in
Eq.(\ref{pot-met}) as
\begin{eqnarray}
\frac{dM(r_+)}{dr_+}&=&\frac{M_+}{r_+}\left(\frac{4M^2_+r_+-4M_+Q^2+Q^2r_+}{4M^2_+r_++4M_+Q^2-Q^2r_+}\right),\label{mass1}\\
\frac{dM(\phi_+)}{d\phi_+}&=&\frac{M_+}{2\phi_+}\left(\frac{4M^2_+\sqrt{\phi_+}
                            -4M_+Q^2+Q^2\sqrt{\phi_+}}
                            {4M^2_+\sqrt{\phi_+}+4M_+Q^2-Q^2\sqrt{\phi_+}}\right),\label{mass2}
\end{eqnarray}
respectively. On the other hand, the derivatives of the
temperature functions with $r_+$ ($\phi$) can be also obtained as
\begin{eqnarray}
\frac{dT(r_+)}{dr_+}&=&\frac{1}{4\pi}
             \left[-\frac{1}{r^2_+}\left(1+\frac{Q^2}{4M^2_+}\right)+\frac{2Q^2}{M_+r^3_+}+
             \left(\frac{Q^2}{M^2_+r^2_+}-\frac{Q^2}{2M^3_+r_+}\right)\frac{dM(r_+)}{dr_+}\right]
             \label{temp1}\\
\frac{dT(\phi_+)}{d\phi_+}&=&-\frac{1}{4\phi^{3/2}_+}+\frac{Q^2}{4M_+\phi^2_+}
                              -\frac{Q^2}{16M^2_+\phi^{3/2}_+}
                    + \left(\frac{Q^2}{4M^2_+\phi_+}-\frac{Q^2}{4M^3_+\sqrt{\phi_+}}\right)
                    \frac{dM(\phi_+)}{d\phi_+}.\nonumber\\
                        \label{temp2}
\end{eqnarray}
Note in these calculations that one should be careful to
differentiate the temperatures with $r_+$ ($\phi$) because they
also have the derivatives of the mass functions as shown in Eqs.
(\ref{mass1}) and (\ref{mass2}). This contrasts to the usual
calculations for the specific heats of the non-linear Born-Infeld
and the RN black holes in which cases the mass functions can be
explicitly separated with the horizon radius, while it is not for
our non-linear MCRBH. Now, combining these equations (\ref{temp1})
and (\ref{temp2}) with (\ref{mass1}) and (\ref{mass2}),
respectively, we have the final expressions of the specific heat,
Eqs.(\ref{2eq6}) and (\ref{vp}), which blow up at the radius $r_m$
($\phi_m$) of giving the maximum Hawking temperature as expected.

\end{document}